\documentclass[12pt]{iopart}

\usepackage{graphicx}
\usepackage{epstopdf}
\usepackage[backend=biber,natbib=true, citestyle=numeric-comp,sorting=none,defernumbers=true,bibencoding=utf8,style=numeric]{biblatex}

\usepackage{color}
\usepackage[useregional]{datetime2}
\begin{document}

\title[Enhancement of quantum capacitance...]{Enhancement of quantum capacitance in graphene electrodes by chemical modifications using aliphatic/aromatic molecules and their radicals }
\author{Sruthi T \& Kartick Tarafder*}
\address{Department of Physics, National Institute of Technology, Srinivasnagar, Surathkal, Mangalore Karnataka-575025, India}
\ead{karticktarafder@gmail.com}
\vspace{10pt}
\begin{indented}
\item[]\today
\end{indented}
\begin{abstract}

We have carried out the systematic study of quantum capacitance (C$_Q$) in functionalized graphene. The functionlization of graphene has been done by doping with different aliphatic and aromatic molecules and their radicals. Using density functional theory calculations, we first analyze the electronic band structure of the functionalized graphene and subsequently obtained the quantum capacitance in each system. We observed that the quantum capacitance can be enhanced by doping aliphatic and aromatic molecules and their radicals on  graphene sheet, especially radical functionalized graphene shows significant enhancement in C$_Q$. Our theoretical investigation reveals that  aromatic and aliphatic radicals generates localized density of states near the Fermi level, due to a charge localization. As a result, a very high quantum  capacitance (above 200 $\mu F/cm^2$) compared to pristine graphene has been estimated. Effects of atomic dislocation and the vacancy defect on graphene during functionalization has also been incorporated in our investigation. Obtain results predict that the  formation of molecular radicals while functionalization of graphene could be an efficient way to synthesize highly efficient graphene based supercapacitor electrode materials which in turn may significantly improve the performance of supercapacitors. 

\end{abstract}

\section{Introduction}

Large scale energy production considering low emission of greenhouse gases have attracted with great research interest due to the ever-rising energy demands. Harvesting of renewable from sun light along with an efficient energy storage technique would be the best alternative in the current scenario. Supercapacitors or ultra-capacitors plays a pivotal role in storing and transporting of harvested energy. They exhibit high power density, faster charging and discharging capabilities and long cycle life with a low maintenance cost.
Two dimensional materials like graphene and their derivatives could play an important role to design an efficient supercapacitor electrodes due to a very high surface volume ratio and its unique electronic behavior\cite{1,2,3}. However, the quantum capacitance of pristine graphene electrode is very small which is one of the major drawbacks. Sufficiently large quantum capacitance of the electrode material is very much essential for obtaining a high energy density.
Chemically functionalized graphene reveals numerous possibilities could be an alternative to overcome this drawback, as the electronic behavior of graphene can be change in a controllable manner\cite{4}. Recent experimental studies have shown that the attachment of inorganic clusters or organic molecules to pristine graphene modifies its electronic properties which is favorable for an electrode material in supercapacitors\cite{5,6}.

Lazar {\sl et. al}.\cite{7} have recently reported the adsorption enthalpies of seven organic molecules on graphene and have estimated the strength of interaction between graphene and the organic molecules through their combined experimental and theoretical study. Other experimental techniques have also been reported to modify the graphene structure upon fuctionalization using small molecules, fragments and their radicals. Lonkar {\sl et al}\cite{8} proposed an efficient functionalization technique using both chemical and physical methods to improve the stability of the modified graphene. They have used both covalent and noncovalent chemical modification of graphene aiming to its  applications in electronic devices and supercapacitors\cite{9,10}.

The recent studies indicate that the quantum capacitance in graphene based electrodes can be improved by covalent as well as noncovalent functionalization using fragments\cite{11,12}. Although many experimental studies have been carried out on functionalized graphene in this direction, but a detail theoretical understanding of modified electronic behavior in fragment functionalized graphene considering various aliphatic and aromatic dopants and their effect on the $C_Q$ is lacking. In this work, we have examined the quantum capacitance of various fragment functionalized graphene using density functional theory calculations. The chemical modifications have been done using different fragmental groups such as alkene, alkyne, ketones, amines, amides, nitriles, carboxilic acids, sulphoxides and aromatics. Since there are possibilities of radical formation of these dopand molecules during the synthesis process, we have also scrutinized the adsorption mechanism of aromatic and aliphatic radicals on graphene surface, their consequence on graphene electronic structure and in quantum capacitance\cite{13,14}. Since formation of defaces and dislocation at the graphene surface is inevitable during functionalization, we have extended our study of quantum capacitance calculation in the fragment-functionalized graphene considering vacancy defects on the graphene surface.

\section{Computational Method} 

In order to accommodate molecules and fragments on the graphene surface we generate large supercells of graphene unit cell and place one fragment in the supercell. A sufficiently large vacuum was considered in the perpendicular direction of graphene sheet (height$\textgreater$15\AA) to avoid the interaction with its periodic images. Depending on fragment's dimension and to avoid fragment-fragment interactions, different size supercells from 4$\times$4$\times$1 (32C atoms) to 8$\times$8$\times$1 (128C atoms) were used so that the distance between two fragments remain more than 5\AA . The vacancy defected configurations were realized by removing C atoms from the graphene sheet on the same supercell. In order to model radical adsorbed graphene surface, we have removed one hydrogen atom from each fragment to make radicals, ie. we removed H from the ring in benzene, from NH$_2$ group in aniline, from CH$_3$ group in toulene, from OH group in phenol, from alpha/beta position in napthalene, from the alpha position in anthracene etc and placed in the unit cell.

To study the electronic structure and quantum capacitance in functionalized graphene, we employed density functional theory (DFT) calculations using Vienna Ab-initio Simulation Package(VASP)\cite{15,16} Projector augmented plane wave (PAW) method was used in each of our simulation process\cite{17}. Generalized Gradient Approximation(GGA) with Perdew-Burke-Ernzerhof(PBE) parameterization were used to approximate the exchange correlation potential\cite{18}. A sufficiently large plane-wave kinetic energy cut-off (\textgreater 400eV) was used in all our calculations for the desired accuracy. A 6$\times$6$\times$1 $\Gamma$ centered k-point mesh was used for reciprocal space sampling in geometry optimization. We have considered 10$^{-6}$H tolerance in total energy in electronic selfconsistency. Inter-atomic forces were obtained using Hellmann-Feynman theorem and subsequently reduced by moving atoms using conjugate gradient algorithm to obtain the minimum energy structure. The process continued untill all inter-atomic forces reduced to 0.01 eV/\AA. Finally, a denser 24$\times$24$\times$1 k-point grid was used for the precise extraction of electronic structure information such as electron density of states (DOS) and atom projected density of states(PDOS) from the optimized geometries.  

In general, the quantum capacitance of materials is defined as 

\begin{equation}
C{_Q}$ = $\frac{dQ}{d\phi}
\end{equation}

where Q is excessive charge on graphene electrode and $\phi$ is the chemical potential\cite{19}. When the chemical potential of electrode is shifted by applying $\phi/e$ potential difference, the excessive charge on the electrode (Q) can be obtain from

\begin{equation}
Q = e\int_{-\infty}^{+\infty} D(E)[f(E) - f(E - \phi)] dE
\end{equation}

Here {\sl e} is the electrons charge (1.6*10$^{-19}$ C),  $\phi$ is the chemical potential, {\sl D(E)} is the DOS, f(E) is Fermi-Dirac distribution function, E refers to Energy with respect to Fermi level and $k$ is the Boltzmann constant. Therefore, in presence of an applied (electrical) potential, one can estimate the excess charge stored on the electrode  material directly form its density of states.

By differentiating Q with respect to $\phi$, we can obtain the expression for quantum capacitance as, 

\begin{equation}
C{_Q} =\frac{dQ}{d\phi} = \frac{e^2}{4kT}\int_{-\infty}^{+\infty} D(E) Sech^{2}\left(\frac{E - {e\phi}}{2kT}\right) dE
\label{QC1}
\end{equation}

Therefore, when the density of states (DOS) is known, the quantum capacitance $C_Q$ of a channel at a finite temperature T can be calculated from equation (\ref{QC1}).  

\section{Results and discussion}

The expression of $C_Q$ in equation (\ref{QC1}) clearly shows that the quantum capacitance of a material is directly proportional to the density of states present near the Fermi energy. For pristine graphene the DOS near Fermi-energy is negligible. However, an efficient functionalization of graphene may tune the density of states near the Fermi level\cite{20,21,22}. The main purpose of functionalizing pristine graphene sheets with organic functional groups is the dispersibility of graphene in common organic solvents which is one of the essential requirements for making a supercapacitor. 
In this present work we model our system considering non-covalent functionalization of graphene with different fragments of aliphatic and aromatic donor and acceptor molecules. In aliphatic group we have considered fragments such as alkene, alkyne, ketones, amines, amides, nitriles, carboxilic acids, sulphoxides. In one of our previous study we showed that the adsorption of N and S atoms on graphene have greater impact on the $C_Q$\cite{23}, therefore amine and sulphoxide groups were selected as dopants which contains N and S adatoms. The piperidine ((CH$_2$)$_5$NH) and dimethyl sulfoxide (DMSO,(CH$_3$)$_2$SO) are selected as amine and sulfoxide groups for functionalization.
In aromatic group, benzene, aniline, phenol, anthracene, toluene and naphthalene were used for functionalization.

 We have carried out our investigation in three different steps. In the first step we investigated the adsorption properties of these molecule and their radicals. 
The stability of the functionalized structure was examined by estimating average adsorption energies $E_{ad}$ of fragments using 

\begin{equation}
E_{ad} = \frac{1}{n}[E_{tot} - E_{gr} -nE_{frag}]
\end{equation}

where $E_{tot}$ is the total energy of the functionalized graphene unit cell,   $E_{gr}$ is the total energy of pristine graphene in the same unit cell, $E_{frag}$ is the per molecule energy of the fragment and $n$ represents the number of adsorbed molecules present in the unit cell. In the second step we have extracted precise electronic structure of the molecule functionalized graphene and finally estimated the quantum capacitance in each system using equation(3). 

\subsection{Aliphatic Molecules and Radical functionalization on pristine graphene}
\begin{figure}[!ht]
	\centering
	\includegraphics[width=0.6\linewidth]{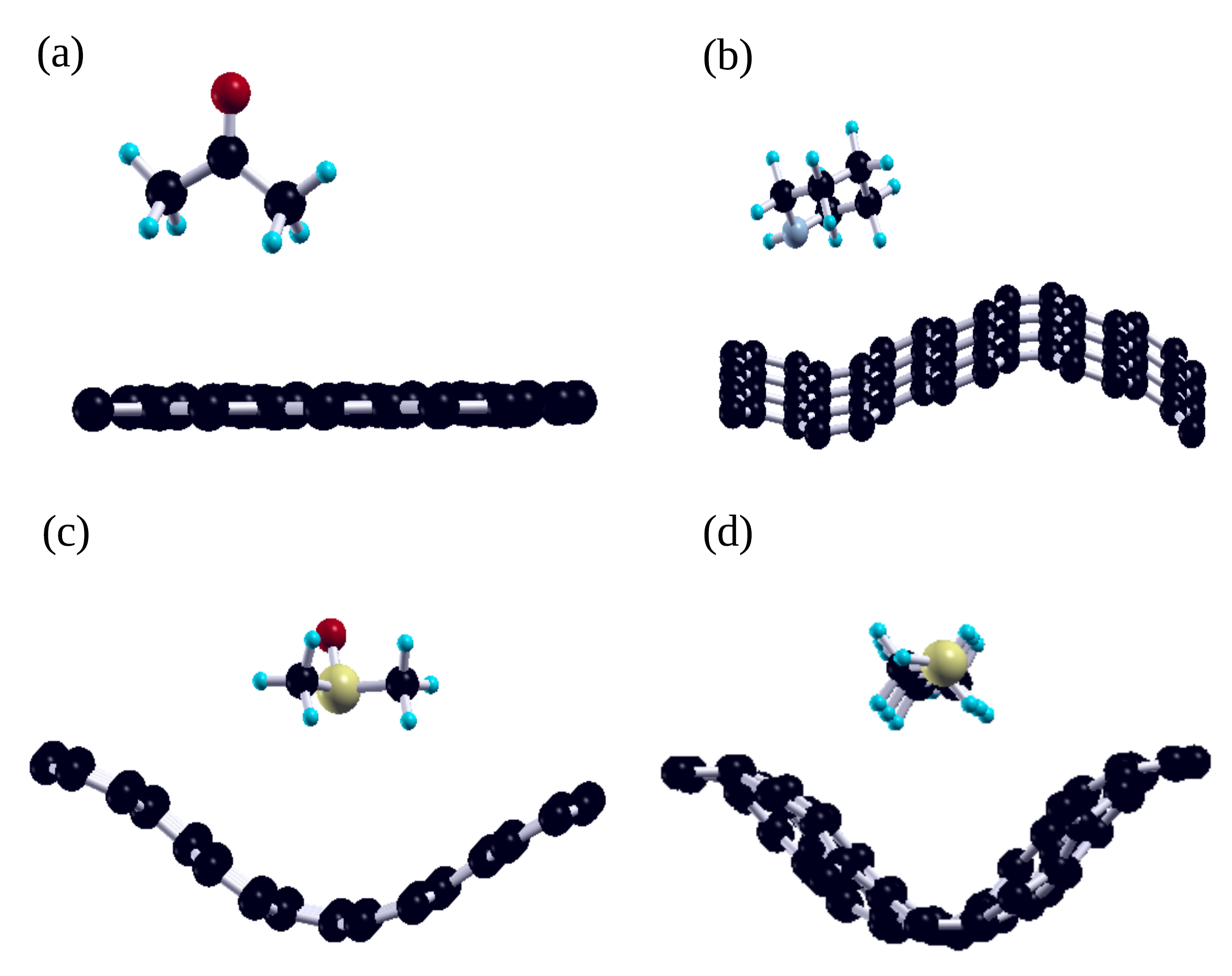}\quad
	\caption{(colour online) Energy-optimized geometry of functionalized graphene with (a)Acetone(ketone), (b)piperidine(amine), (c)DMSO and (d)Hexane-1-thiol. Black, grey, blue, red and yellow ball represents C, N, H, O and S atoms respectively.}
	\label{1}
\end{figure}

Aliphatic molecules are strongly interacting with graphene surface. We observed significant distortion of the graphene structure upon adsorption of acetamide, piperidine, DMSO, butironitrile and hexane-1-thiol, where as the distortion is negligible for alkene, acetic acid and ketone adsorption. In case of large deformation of structure, the total adsorption energy for these molecules $E_{ad}$ includes a large energy cost to deform the graphene lattice, which can be calculated from E($\bar{g}$)-E(g), where E($\bar{g}$) is the energy of the deformed graphene. We have excluded the deformation energy from the total adsorption energy and listed in the Table.[\ref{table1}]. Calculated adsorption energies for different aliphatic molecules on graphene indicate that most of the aliphatic molecule considered in this study are well absorb on the graphene surface. The energy-optimized acetone(ketone), piperidine(amine groups), DMSO and Hexane-1-thiol functionalized graphene are shown in Fig.[\ref{1}]. 

\begin{figure}[!ht]
	\centering
	\includegraphics[width=1.1\linewidth]{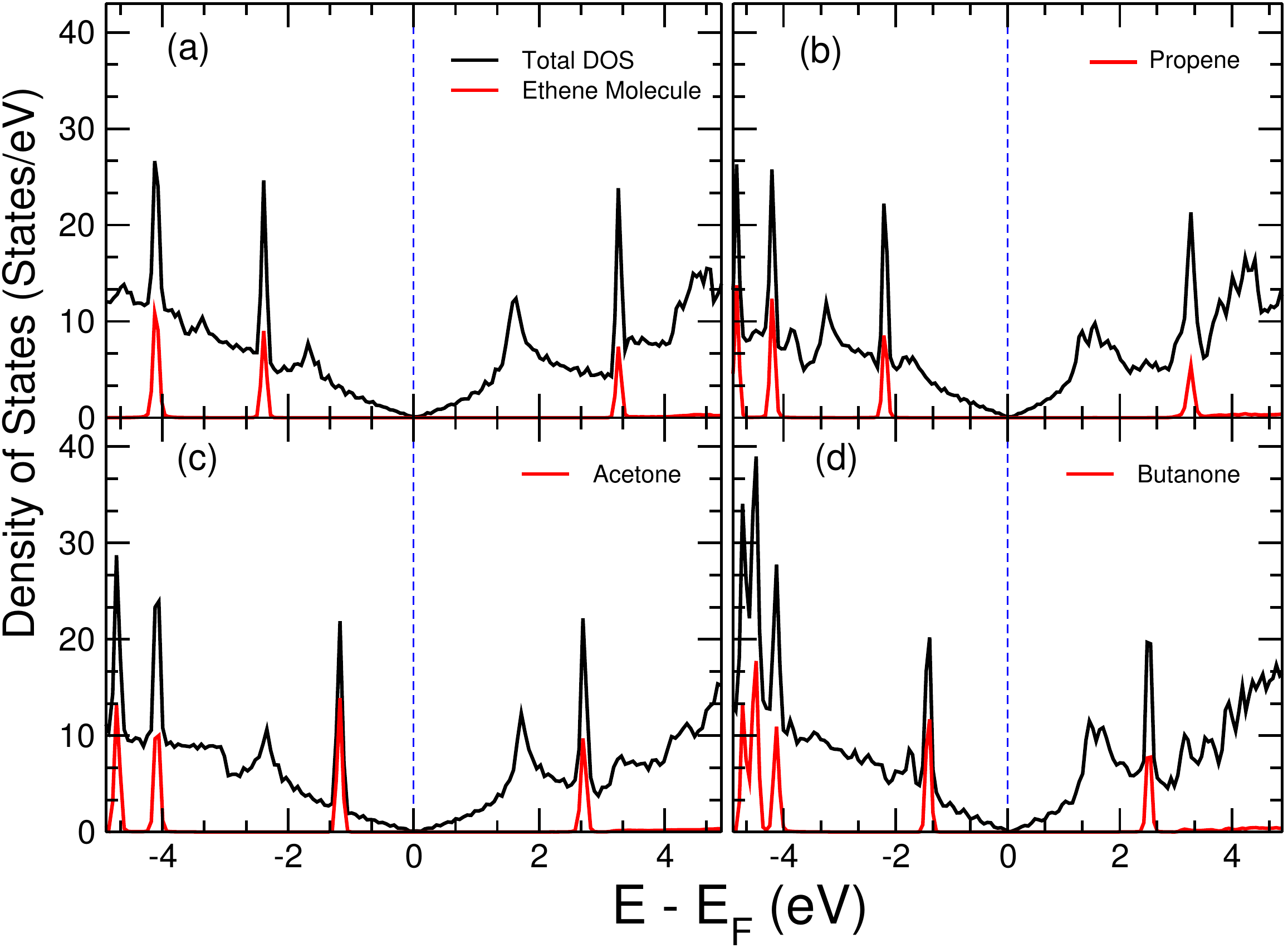}\quad
	\caption{(colour online) Projected density of states for functionalized graphene with (a)Acetone(ketone), (b)piperidine(amine), (c)DMSO (sulfoxide) and (d)Hexane-1-thiol. The colored curve represents dos from the aliphatic molecules. The vertical blue dashed line is the Fermi energy set at E=0. }
	\label{2}
\end{figure}

Although a significant geometric distortions were observed in many aliphatic functionalized graphene, but the electronic structure of the graphene is not much perturbed. Atom projected density of states for ketone(acetone), amine(piperidine), sulfoxide(DMSO) and Hexane-1-thiol functionalized graphene are shown in Fig.[\ref{2}a-d]. Clearly the Dirac cone structure is preserved in each case. The density of states contributed from aliphatic molecules in the system are highly localized and located in the valence band far from the fermi level. Since the conduction in pristine graphene is mainly due to the de-localized $\pi$ cloud from p$_z$ orbitals of carbon, only C p-states are  present around the Fermi-energy in the functionalized system. Surprisingly strong adsorption of these molecule on graphene does not produce significant change in the electronic structure. Only noticeable change is observed in case of thiol adsorption, but DOS localization near fermi energy is negligible. Bader charge analysis shows that a small charge transfer occur between graphene and the fragments as shown in the Table.[\ref{table1}]. Fragmental groups such as butiro nitrile and hexane-1-thiol behave as electron acceptors on graphene and ethene, acetylene, acetamide, piperidine, acetic-acid, acetone, DMSO are behaving as electron donor in functionalized graphene system. This clearly indicates that the interaction between aliphatic molecule and graphene is mainly Van der-Waals type. The overall change in density of states near the Fermi energy is very small for each aliphatic molecule functionalized graphene results to a small change in quantum capacitance compared to pristine graphene tabulated in Table.[\ref{table1}]. Atomic defect on pristine graphene causes charge localization, which inturn may improve the quantum capacitance in functionalized graphene. However we have not observed any significant change in quantum capacitance after introducing vacancy defects in aliphatic functionalized graphene.

\begin{table}[!ht]
	\caption{\label{table1} Calculated $E_{ad}$ value, charge transferred and quantum capacitance value at 300 K for various Aliphatic molecules functionalized graphene with single molecule doping.}
	\begin{indented}
		\item[]\begin{tabular}{@{}llll}
			\br
			\textbf {Configuration} & \textbf{ Adsorption }& \textbf{charge }
			& \textbf{Quantum} \\
			& \textbf{energy(eV)} & \textbf{transferred(e)}   & \textbf{capacitance ($\mu F/cm^2$)}              \\
			\mr
			Pristine Graphene   & -              & -         & 1.3000 \\
			FG-Ethene     &  -0.24157376   &  0.120887 & 1.9323  \\
			FG-Acetylene &  -1.175215605  &  0.216626 & 4.4936  \\
			FG-Acetamide &  -1.05744605   &  0.909416 & 3.6581  \\
			FG-Piperidine   &  -1.14756113   &  0.046263 & 3.7188  \\
			FG-Acetic-Acid   &  -0.573780565  &  0.144073 & 2.5325  \\
			FG-Acetone   &  -1.68168973   &  0.124616 & 1.7116  \\
			FG-Butiro-Nitrile   &  -1.17548787   & -0.077611 & 3.4606  \\
			FG-DMSO   &  -0.59555937   &  0.142819 & 3.5130 \\
			FG-Hexane-1-Thiol   &  -2.26807395   & -0.055368 & 8.1710   \\
			\br
		\end{tabular}
	\end{indented}
\end{table}

\begin{figure}[!ht]
	\centering
	\includegraphics[width=0.9\linewidth]{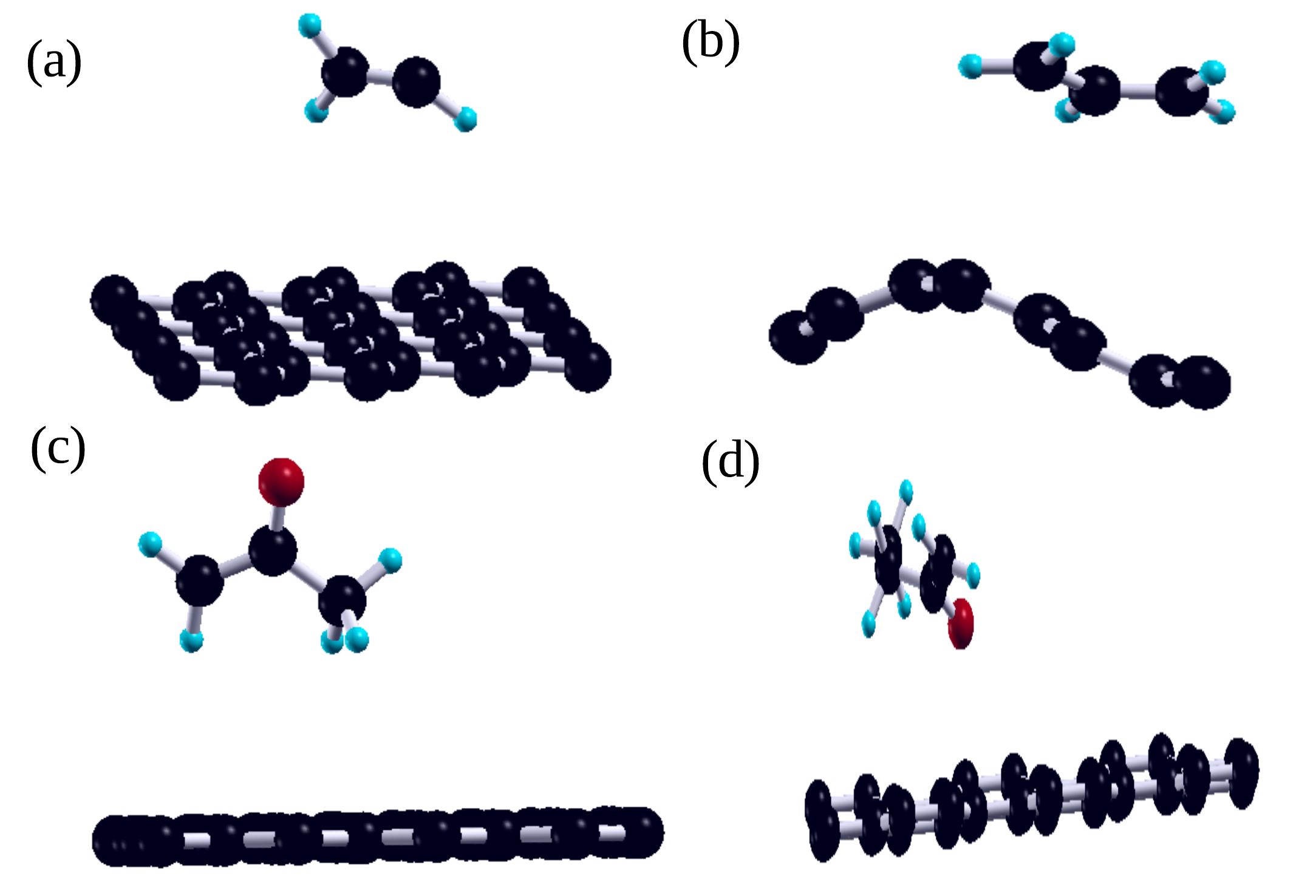}\quad
	\caption{(colour online) Energy-optimized geometry of (a)Ethene, (b)Propene, (c)Acetone and (d)Butanone radical functionalized graphene. Black, blue, and red ball represents C, H and O atoms respectively.}
	\label{alip-rad-geo}
\end{figure}

Since localized states are found in molecular radicals, therefore radical functionlization may help to create localized state near the fermi energy of the system. We next investigated the electronic structure and quantum capacitance in aliphatic radical functionalized graphene. We have created radicals out of ethene, propene, acetone and butanone molecules by removing one of the hydrogen atoms from the fragments. The adsorption energies and charge transferred for different aliphatic radical functionalized graphene are listed in the Table.[\ref{table2}]. Negative values of the adsorption energy confirm that all aliphatic radicals are stable on top of graphene sheet. A notable distortion in graphene sheet was observed, in which graphene carbon atoms (near radicals) are attracted towards radicals. Optimized geometry of ethene, propene, acetone and butanone radical functionalized graphenes are shown in Fig.[\ref{alip-rad-geo}]

\begin{figure}[!ht]
	\centering
	\includegraphics[width=1.1\linewidth]{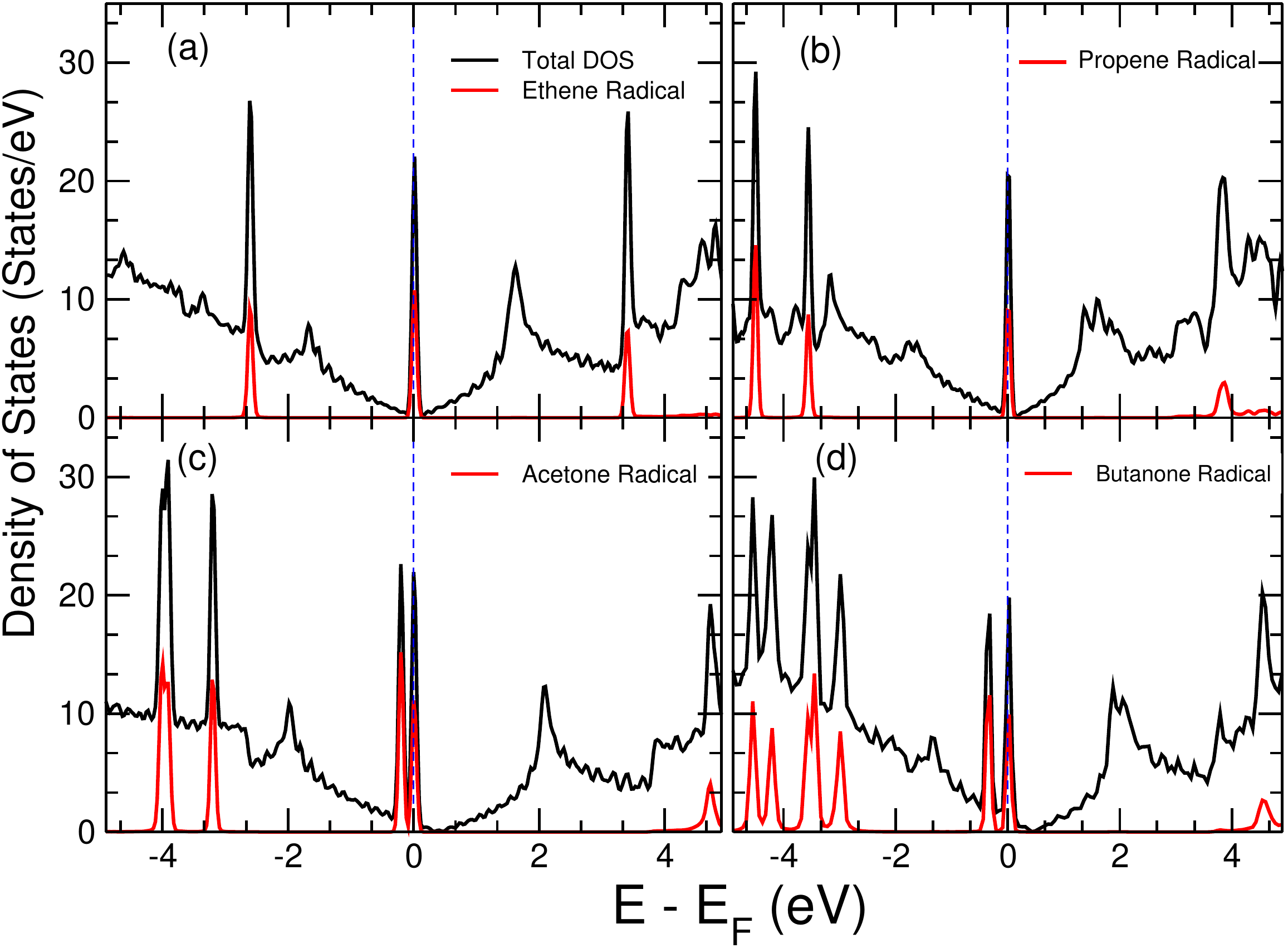}\quad
	\caption{(colour online) Projected density of states for functionalized graphene with (a)Ethene, (b)Propene, (c)Acetone and (d)Butanone radicals. The colored curve represents dos from the aliphatic radical. The vertical blue dashed line is the Fermi energy set at E=0.}
	\label{4}
\end{figure}

We observed  significant changes in electronic behavior of aliphatic radical doped graphene. The Dirac cone structure got distorted. From the density of states calculation we observed that the valence band maximum is contributed mainly from the atoms of aliphatic radicals in each case and the DOS near Fermi level is very high. For acetone radical functionalized graphene accumulation of density of state near the Fermi energy is maximum as shown in Fig.[\ref{4}c], as a result high quantum capacitance value of 234 $\mu F/cm^2$ is observed in this system. 

 In case of other aliphatic radical-functionalized graphene, the density of states calculation also shows a similar behavior. A strong peak near $E_F$ appear for each of ethene, propene, and butanone functionalized graphene which are mainly contributed from atoms of the radicals as shown in Fig.[\ref{4} (a),(b) \& (d)] respectively. The calculated values of quantum capacitance for these systems are very high (in the range of 200 $\mu F/cm^2$) listed in Table.[\ref{table2}]. 
 
\begin{table}[!ht]
	\caption{\label{table2} Calculated Adsorption value, charge transferred and $C_Q$ value at 300 K for various aliphatic radical functionalized graphene with single radical doping.}
	\begin{indented}
		\item[]\begin{tabular}{@{}llll}
			\br
			\textbf {Configuration} & \textbf{ Adsorption }& \textbf{charge } & \textbf{Quantum} \\
			& \textbf{energy(eV)m} & \textbf{transferred(e)} & \textbf{capacitance ($\mu F/cm^2$)}            \\ 
			\mr
			Pristine Graphene  		& - & - & \ \ \  1.3000 \\
			G-Ethene    			&  -0.27028465  &  -0.011299  &  220.6126  \\ 
			G-Propene				&  -0.31086963  &  -0.016822  &  221.2949   \\ 
			G-Acetone   			&  -1.81821633  &  -0.083088  &  234.6769  \\
			G-Butanone  		    &  -0.93227212  &  -0.108354  &  211.4947  \\
			\br
		\end{tabular}
	\end{indented}
\end{table}

\subsection{Aromatic molecules and Radical functionalization on pristine graphene}

\begin{figure}[!ht]
\centering
\includegraphics[width=1\linewidth]{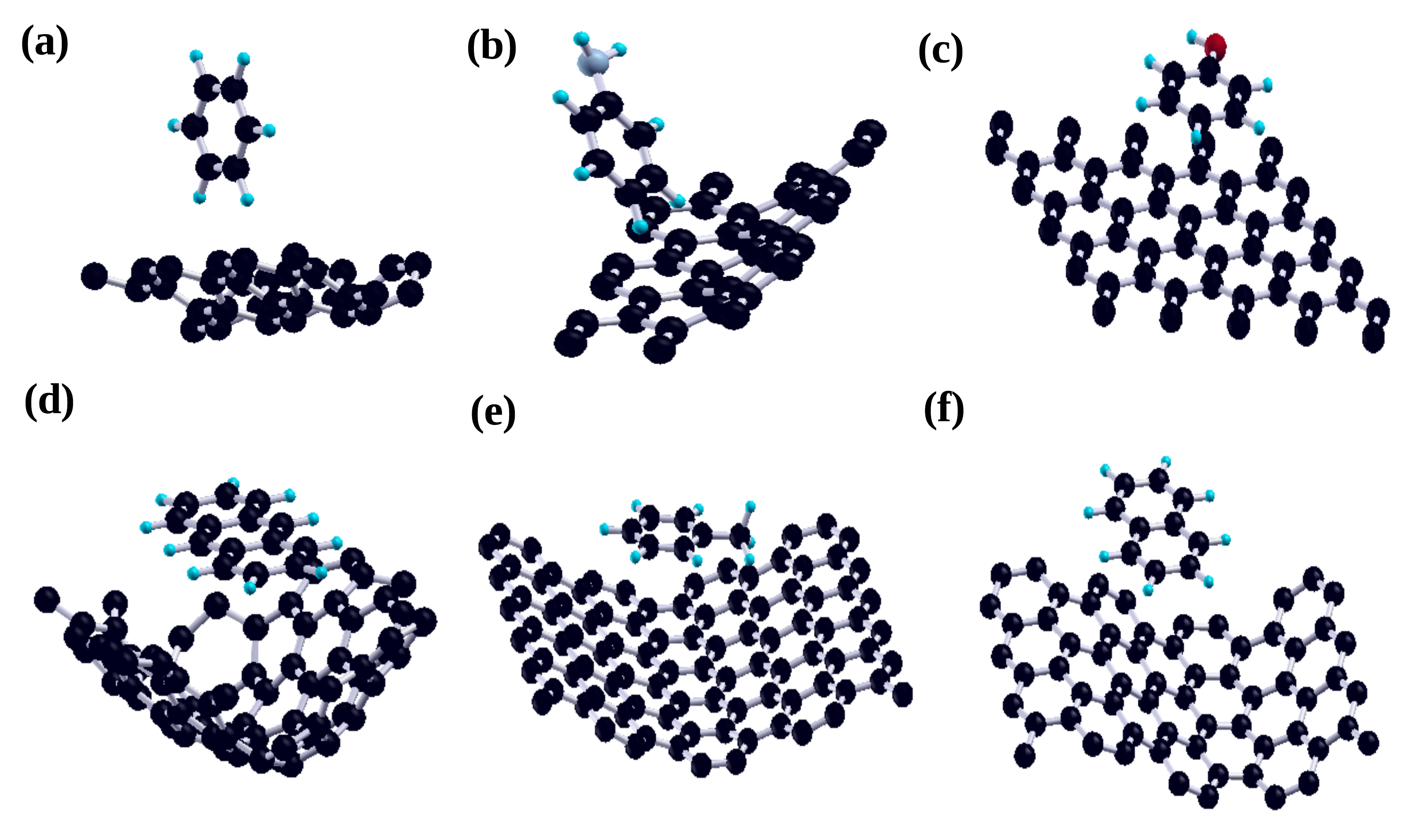}\quad
\caption{(colour online) Energy-optimized geometry of (a)Benzene, (b)Aniline, (c)Phenol, (d)Anthracene, (e)Toluene and (d)Naphthalene molecule functionalized graphene. Black, blue, red, and grey ball represents C, H, O and N atoms respectively.}	
\label{aro-gr-str}
\end{figure}

\begin{figure}[!ht]
	\centering
	\includegraphics[width=1.1\linewidth]{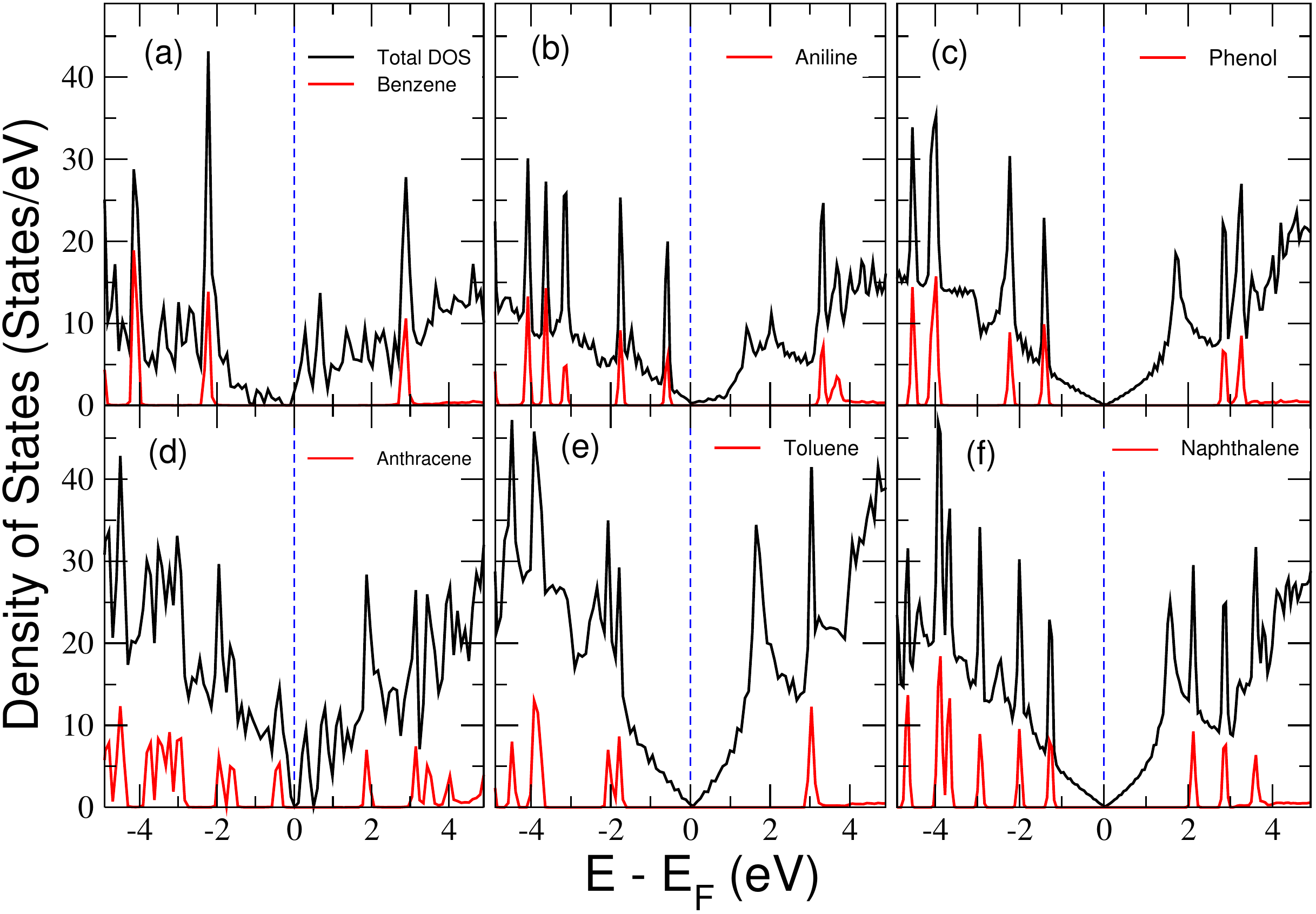}\quad
	\caption{(colour online) Projected density of states for functionalized graphene with (a)Benzene, (b)Aniline, (c)Phenol, (d)Anthracene, (e)Toluene and (d)Naphthalene molecule. The colored curve represents dos from the aromatic molecules. The vertical blue dashed line is the Fermi energy set at E=0. }
	\label{aro-gr-dos}
\end{figure}

We have used benzene, aniline, phenol, anthracene, toluene and naphthalene molecules and their radicals to functionalize graphene and study the electronic structure and subsequently the change in quantum capacitance. The stable adsorption position of aromatics on graphene was calculated by placing molecules in different possible configurations and comparing the total energies. The perpendicular configuration (ie molecular plane is perpendicular to the graphene surface)was found to be most favorable for benzene and aniline molecule; where as a parallel configuration, ie molecular plane is parallel to the graphene surface was found to be the most stable configuration for phenol, anthracene, toluene and naphthalene. 
 Interestingly, atom of molecules which is closest to the graphene surface in  perpendicular configurations is sitting at the center of graphene hexagonal ring.  One reason is that this configuration  minimizes the repulsion between electrons in $\pi$-orbitals and maximizing the attractive interaction between the molecule and the graphene. Our calculation shows physisorption for all these molecules on graphene surface with a distance of $\sim$3.0\AA  \ from the graphene surface. Optimized geometric structures of six aromatic functionalized graphene structures are shown in Fig.[\ref{aro-gr-str}].
Out of six aromatic molecules only benzene, aniline and anthracene functionalization shows a noticeable change in electronic structure of graphene near the Fermi energy. The calculated electronic density of states for six aromatic molecule functionalized graphene are shown in Fig.[\ref{aro-gr-dos}]. Although the band gap in these three systems are zero, but the linear dispersion is missing in the band structure. Density of states are localized near the Fermi energy which are mainly contributed from the graphene itself.  
 Bader charge analysis shows that in case of toluene and anthracene the charge transfer is significant. We notice that the toluene and aniline acts as electron donor and anthracene as electron acceptor. The adsorption energies and charge transferred for different aromatic molecules on graphene sheet are listed in the Table.[\ref{table3}]. There are no significant improvement  in the  value of quantum capacitance was found in these case. The maximum value observed is 29.09 $\mu F/cm^2$ for benzene functionalized graphene.  C$_Q$ values for other aromatic functionalized graphenes  are  shown in Table.[\ref{table3}].

\begin{table}
	\caption{\label{table3} Calculated $E_{ad}$ value, charge transferred and quantum capacitance value at 300 K for various aromatics functionalized graphene with single molecule doping.}
	\begin{indented}
		\item[]\begin{tabular}{@{}llll}
			\br
			\textbf {Configuration} & \textbf{ Adsorption }& \textbf{charge } & \textbf{Quantum} \\
			& \textbf{energy(eV)m} & \textbf{transferred(e)} & \textbf{capacitance ($\mu F/cm^2$)}            \\ 
			\mr
			Pristine Graphene  		&  - & - &  \ \ 1.30 \\
			G-Benzene    			&  -2.28560334   & -0.062481  & 29.090    \\ 
			G-Aniline				&  -1.31072759   &  0.062364  & 6.2489  \\ 
			G-Phenol			 	&  -1.18273005   & -0.039488  & 2.4905 \\
			G-Anthracene  			&  -1.32664995	 & -0.373914  & 4.4059 \\
			G-Toluene 		    	&  -1.21070580   &  0.326792  & 6.3420 \\
			G-Naphthalene  		    &  -0.70095142   & -0.013112  & 3.4066 \\
			\br
		\end{tabular}
	\end{indented}
\end{table}

In order to enhance the Q$_C$ value we introduced vacancy defects in pristine graphene. The distortion in the graphene sheet is more in defected graphene compared to the pristine graphene. In this case, the adsorbed molecule come close to the defect site and bonded to the carbon atom of graphene. This  gives rise to strong hybridization between  $\pi$  orbitals of graphene and molecule, as a result the density of states present near the Fermi level get reduced leads to a smaller value of quantum capacitance. We observed no significant increment in quantum capacitance for aromatic molecules functionalized on top of vacancy defected graphene.

\begin{figure}[!ht]
\centering
\includegraphics[width=1\linewidth]{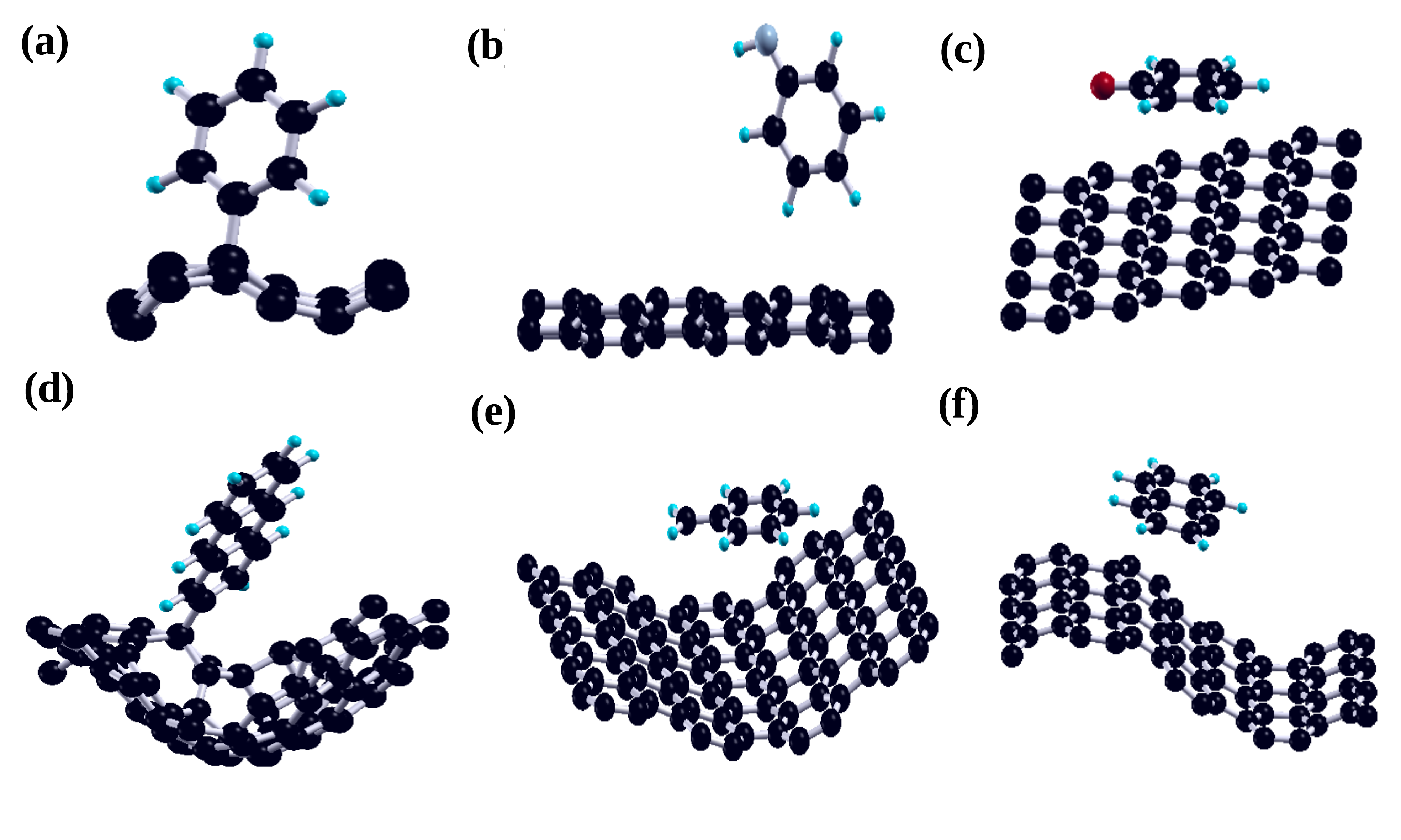}\quad
\caption{(colour online) Energy-optimized geometry of (a)Benzene, (b)Aniline, (c)Phenol, (d)Anthracene, (e)Toluene and (d)Naphthalene radical functionalized graphene. Black, blue, red, and grey ball represents C, H, O and N atoms respectively. }
	\label{aro-rad-gr-str}
\end{figure}

To understand the effect of aromatic radical functionalization of graphene on quantum capacitance, radicals were formed by removing one H atom from different position of the molecule and placed in the same supercell which was used in molecule functionalization. Geometry optimization shows different adsorption configurations, in case of benzene and anthracene radicals, radical sites are coming closer to the graphene and bonded with graphene surface, where as phenol toluene and naphthalene radicals are sitting parallel to the surface. In the case of aniline radicals, it is  stabilizing in such a way that the unsaturated atom of the molecule keeps maximum distance from the graphene surface. The optimized geometry of aromatic radical adsorbed graphene are given in Fig.[\ref{aro-rad-gr-str}]. 

Calculated adsorption energies of aromatic radicals on graphene surface are more compare to the respective molecule absorptions.   

\begin{figure}[!ht]
	\centering
	\includegraphics[width=1\linewidth]{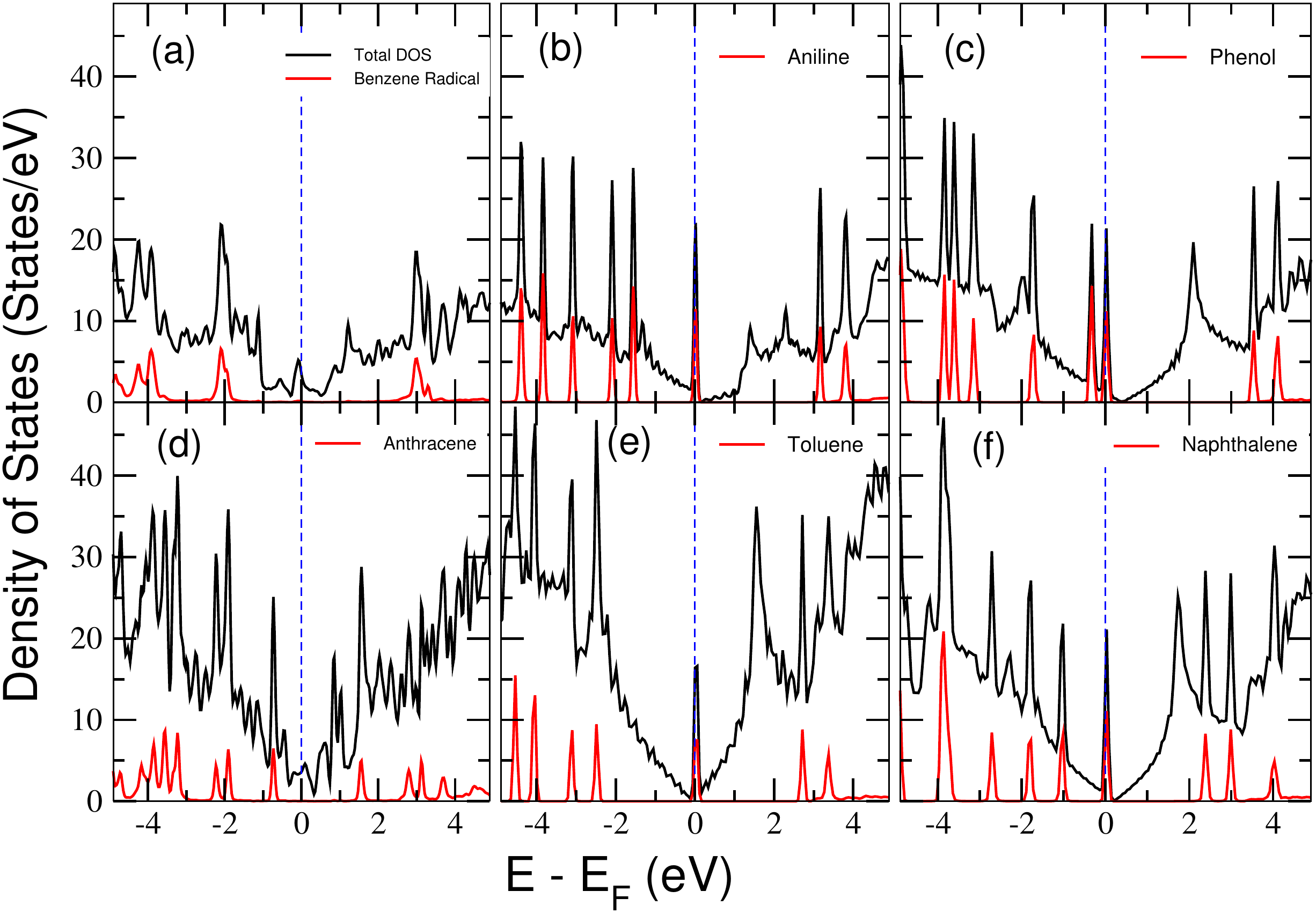}\quad
	\caption{(colour online) Projected density of states for functionalized graphene with (a)Benzene, (b)Aniline, (c)Phenol, (d)Anthracene, (e)Toluene and (d)Naphthalene radical. The colored curve represents dos from the aromatic radicals. The vertical blue dashed line is the Fermi energy set at E=0.}
	\label{aro-rad-gr-dos}
\end{figure}

Noticeable change has also been observed in the electronic structure of each aromatic radical functionalized graphene. The Dirac cone structure got distorted in all case, and the valence bands become more dense compared to the conduction bands. Maximum change in electronic structure was observed for benzene and anthracene radical functionalized systems that shows a strong peak near E$_F$ which are 3p$_z$ states from the graphene carbon atom as shown in Fig.[\ref{aro-rad-gr-dos}]. In case of phenol radical doped graphene, states near the Fermi energy are mainly contributed from the oxygen atoms of the radical. 
Two types of naphthalene radicals can be made from the napthalene molecule by removing H from alpha- and beta-positions. We have considered both the radicals, which shows a very similar change in electronic behavior shown in Fig.[\ref{aro-rad-gr-dos}(f)].

We observe that  the accumulation of large density of state near the Fermi energy for all aromatic radical functionalized graphene and relatively large $C_Q$ was obtained for each system as listed in the table[\ref{table4}]. The maximum value of $C_Q = 236.62 \mu F/cm^2$ obtained for phenol radical doped graphene. 
 
\begin{table}
	\caption{\label{table4} Calculated $E_{ad}$ value, charge transferred and quantum capacitance value at 300K for various aromatic radical functionalized graphene with single radical doping.}
	\begin{indented}
		\item[]\begin{tabular}{@{}llll}
			\br
			\textbf {Configuration} & \textbf{ Adsorption }& \textbf{charge } & \textbf{Quantum} \\
			& \textbf{energy(eV)m} & \textbf{transferred(e)} & \textbf{capacitance ($\mu F/cm^2$)}            \\ 
			\mr
			Pristine Graphene  	& - & - & \ 1.3000 \\
			G-Benzene    		&   -4.19110000    & -0.158037      &  56.5226\\ 
			G-Aniline			& 	-2.32103451	   & -0.067172 	    & 226.0405 \\ 
			G-Phenol(para)		&   -1.26131271	   & -0.271752  	& 236.6246  \\
			G-Anthracene   		&   -4.25820133    & -0.461274    	&  59.5798  \\
			G-Toluene  		   	&   -2.54358801    &  0.302454	 	& 201.5426\\
			G-Naphthalene  		&   -0.73822562    & -0.202179	    & 213.5178 \\
			\br
		\end{tabular}
	\end{indented}
\end{table}

To analyze the result we have performed Bader charge analysis on radical functionalized graphene, which shows that there are significant charge transfer between graphene and the aromatic radicals compared to aromatic molecules. In case of anthracene and phenol radical doping, the charge transfer is remarkably high. We found 0.46e and 0.27e charge transfer to anthracene and phenol radicals respectively from the graphene sheet. Where as toluene radical acts as an electron donor, donates +0.3e charge per radical to the graphene sheet. The introduction of the electron accepting/donating radicals disrupts the homogeneity of the charge distribution. Note that, in case of benzene and anthracene the quantum capacitance are relatively low. To understand the result we have calculated the transferred charge densities (from charge densities  before and after adsorption)  for each system.  
\begin{figure}[!ht]
	\centering
	\includegraphics[width=1.0\linewidth]{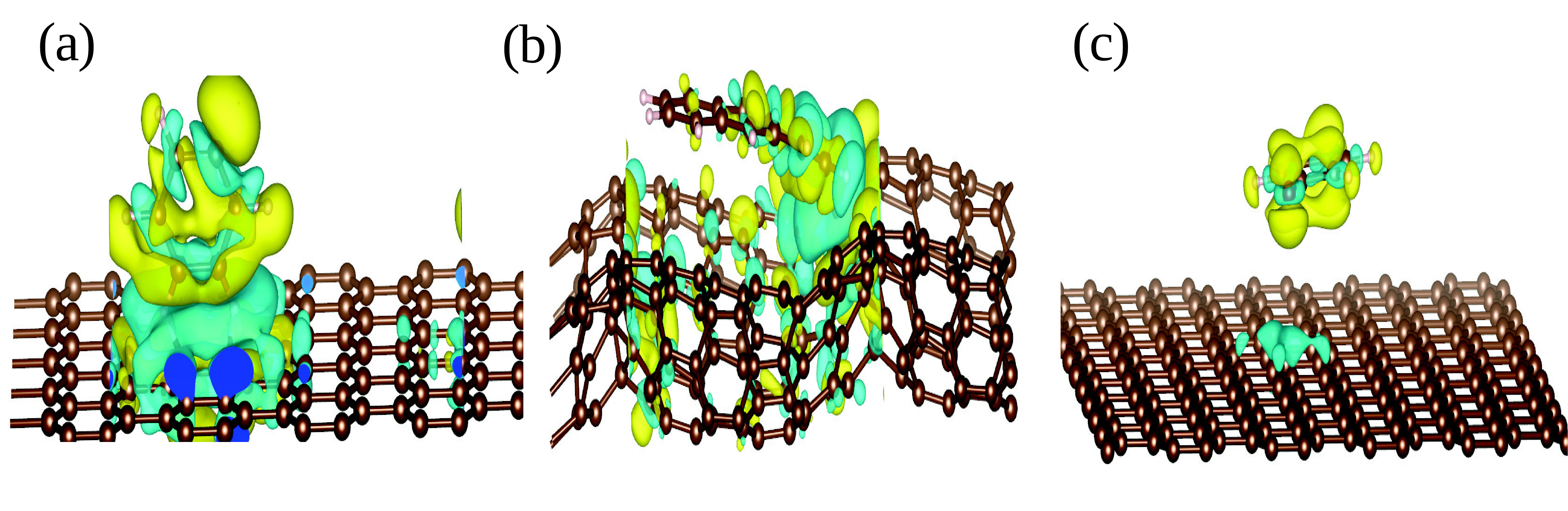}\quad
	\caption{(colour online) Distribution of transferred charge due to (a) benzene (b)anthracene and (c) phenol radical functionalization on graphene. The transferred charges are highly localized in case of phenol but delocalized in benzene and anthracene radical adsorptions}
	\label{charg-dist}
\end{figure}
We observed that   the transfer charges are localized near the unsaturated atoms in case of aniline, phenol, toluene and naphthalene radical doped system. Whereas in case of anthracene and benzene radical doping, the transferred charge is distributed evenly on a comparatively large area. Charge density difference for benzene, anthracene and phenol radical functionalized graphene are shown in Fig.[\ref{charg-dist}]. From the transferred charge density plot one can conclude that the hybridization between graphene and  benzene(or anthracene) radical orbitals are  much more stronger compare to the other radicals adsorptions that pushes localized states lower in energy deep into the valence band as shown in  Fig.[\ref{aro-rad-gr-dos}-a \& d] respectively. Hence the electronic density of states near the Fermi level decreases for these two systems and consequently the quantum capacitance value  reduced to 56.52 $\mu F/cm^2$ \& 59.57 $\mu F/cm^2$ for benzene and anthracene radical functionalization respectively.  

\section{Conclusion}  

In conclusion, density functional theory calculations were performed to analyze the electronic structure of the functionalized graphene doped with various fragments of aliphatic, aromatic molecules and their radicals. The quantum capacitance in these system were obtained subsequently. Our results shows that aromatic/aliphatic molecules are strongly adsorb on the graphene sheets via strong $\pi$-interactions. A significant charge transfer and charge redistribution occur in donor/acceptor molecule doped graphene. The additional charge carriers brought by aromatic/aliphatic molecules and their radicals change the carrier concentration in  graphene that leads to the shift of the Fermi level. Our study also reveals that the quantum capacitance of graphene-based electrodes can be enhanced by functionalizing it with aliphatic, aromatic molecule and their radicals. The enhancement is consequential, when the graphene is functionalized with radicals rather than their respective molecules. These radicals are behaving as impurities in the system, generates localized density of states near the Fermi energy. Consequently the values of quantum capacitance got increased. The maximum value of quantum capacitance ($\sim 237 \ \mu F/cm^2$) is obtained for phenol-radical(aromatic) functionalized graphene. A very similar value ( $\sim 235 \ \mu F/cm^2$) was obtain for acetone-radical (aliphatic) functionalized graphene.

\section*{Acknowledgement}
KT would like to acknowledge NITK-high performance computing facility and also would like thank DST-SERB(project no. SB/FTP/PS-032/2014 ) for the financial support.

\Bibliography{99}

\item Novoselov K. S. et al. (2005) ‘Two-dimensional gas of massless Dirac fermions in graphene’, {\it Nature}, {\bf 438}(7065), pp. 197-200. \\

\item El-Kady M. F., Shao Y. and Kaner, R. B. (2016) ‘Graphene for batteries, supercapacitors and beyond’, {\it Nature Reviews Materials}, {\bf 1}(7), p. 16033. \\

\item Zhang L. L., Zhou R. and Zhao X. S. (2009) ‘Carbon-based materials as supercapacitor electrodes’, {\it Journal of Materials Chemistry}, {\bf 38}(29), pp. 2520-2531. \\

\item Ke Q. and Wang J. (2016) ‘Graphene-based Materials for Supercapacitor Electrodes - A Review’, {\it Journal of Materiomics}. Elsevier Ltd,{\bf 2}(1), pp. 37-54.  \\

\item Kaverzin A. A., Strawbridge S. M. ,  Price A. S.,  Withers F.,  Savchenko A. K., and Horsell D. W.   (2011), 'Electrochemical Doping of Graphene with Toluene.' {\it Carbon. Elsevier Ltd}, {\bf 49}(12), pp. 3829–3834.

\item Roychoudhury, S., Motta, C. and Sanvito, S. (2016) ‘Charge transfer energies of benzene physisorbed on a graphene sheet from constrained density functional theory’, {\it Physical Review B}, {\bf 93}(4), pp. 1–8. 

\item  Petr L.,  Karlick\'{y} F., Jurecka P. , Kocman M., Otyepkov\'{a} E., Safarova K, and  Otyepka M. (2013) 'Adsorption of Small Organic Molecules on Graphene.' {\it Journal of the American Chemical Society}, {\bf 135}(16), pp.6372–77.\\ 

\item Lonkar S. P., Deshmukh Y. S. and Abdala A. A. (2015) ‘Recent advances in chemical modifications of graphene’, {\it Nano Research},  {\bf 8}(4), pp. 1039–1074.

\item Rubio-Pereda P. and Takeuchi N. (2013) ‘Density functional theory study of the organic functionalization of hydrogenated graphene’, {\it Journal of Physical Chemistry C}, {\bf 117}(36), pp. 18738–18745.

\item Georgakilas V.,  et al. (2012), 'Functionalization of Graphene: Covalent and Non-Covalent Approach.” {\it Chemical Reviews}, {\bf 112}(11), pp. 6156–6214.

\item Plachinda P., Evans D. and Solanki R. (2017) ‘Electrical properties of covalently functionalized graphene’, {\it AIMS Materials Science}, {\bf 4}(2), pp. 340–362. 

\item Moraes, E. E. de.  et al.  (2019), 'Density Functional Theory Study of $\pi$-Aromatic Interaction of Benzene, Phenol, Catechol, Dopamine Isolated Dimers and Adsorbed on Graphene Surface.' {\it Journal of Molecular Modeling} {\bf 25}(10).

\item Su, Q., et al. (2009), 'Composites of Craphene with Large Aromatic Molecules.' {\it Advanced Materials} {\bf 21}(31), pp. 3191–95.

\item Lin H., Fratesi G. and Brivio G. P. (2015) ‘Graphene magnetism induced by covalent adsorption of aromatic radicals’, {\it Physical Chemistry Chemical Physics. Royal Society of Chemistry}, {\bf 17}(3), pp. 2210–2215. 

\item Kresse G. and Furthmüller J. (1996) ‘Efficiency of ab-initio total energy calculations for metals and semiconductors using a plane-wave basis set’, {\it Computational Materials Science}, {\bf 6}(1), pp. 15-50. \\

\item Kresse G. and Furthmüller J. (1996) ‘Efficient iterative schemes for ab initio total-energy calculations using a plane-wave basis set’, {\it Phys. Rev. B. American Physical Society},{\bf 54}(16), pp. 11169–11186 \\.

\item Blochl P. E. (1994) ‘Projector augmented-wave method’, {\it Phys. Rev. B. American Physical Society}, {\bf 50}(24), pp. 17953-17979.\\

\item Perdew J. P., Burke K. and Ernzerhof M. (1997) ‘Generalized Gradient Approximation Made Simple’, {\it Phys. Rev. Lett. American Physical Society}, {\bf 78}(7), p. 1396.\\

\item Mousavi-Khoshdel  M., Targholi E. and Momeni M. J. (2015) ‘First-Principles Calculation of Quantum Capacitance of Codoped Graphenes as Supercapacitor Electrodes’, {\it Journal of Physical Chemistry C}, {\bf 119}(47), pp. 26290-26295. \\

\item Pak A. J., Paek E. and Hwang G. S. (2014) ‘Tailoring the performance of graphene-based supercapacitors using topological defects: A theoretical assessment’, {\it Carbon. Elsevier Ltd}, {\bf 68}(512), pp. 734-741.  \\

\item Jeong H. M. et al. (2011) ‘Nitrogen-doped graphene for high-performance ultracapacitors and the importance of nitrogen-doped sites at basal planes’,{\it  Nano Letters}, {\bf 11}(6), pp. 2472–2477. \\

\item Yang G. M. et al. (2015) 'Density Functional Theory Calculations for the Quantum Capacitance Performance of Graphene-Based Electrode Material.' {\it Journal of Physical Chemistry C}, {\bf 119}(12), pp. 6464–70. \\

\item T. Sruthi and Tarafder, K. (2019) 'Route to achieving enhanced quantum capacitance in functionalized graphene based supercapacitor electrodes'{\it J. Phys:Condensed matter},{\bf 31} pp. 475520

\endbib

\end{document}